# Spending is not Easier than Trading: On the Computational Equivalence of Fisher and Arrow-Debreu Equilibria

Xi Chen[1] and Shang-Hua Teng[2]

[1] Princeton University
[2] University of Southern California

**Abstract.** It is a common belief that computing a market equilibrium in Fisher's spending model is easier than computing a market equilibrium in Arrow-Debreu's exchange model. This belief is built on the fact that we have more algorithmic success in Fisher equilibria than Arrow-Debreu equilibria. For example, a Fisher equilibrium in a Leontief market can be found in polynomial time, while it is PPAD-hard to compute an approximate Arrow-Debreu equilibrium in a Leontief market.

In this paper, we show that even when all the utilities are additively separable, piecewise-linear, and concave functions, finding an approximate equilibrium in Fisher's model is complete in PPAD. Our result solves a long-term open question on the complexity of market equilibria. To the best of our knowledge, this is the first PPAD-completeness result for Fisher's model. [3]

## 1 Introduction

### 1.1 Market Equilibria: Fisher's Model vs Arrow-Debreu's Model

In 1891, Irving Fisher introduced one of the most fundamental exchange market models in his Ph.D. thesis [2]. It considers a market in which there are $n$ buyers and $m$ divisible goods. We denote the amount of good $j$, $j \in [m]$, in the market by $c_j > 0$. Every buyer $i$ comes to the market with a certain amount of money, denoted by $w_i > 0$. The goal of a buyer is to obtain a bundle of goods, denoted by $\mathbf{a}_i \in \mathbb{R}_+^m$, that maximizes her utility function $u_i : \mathbb{R}_+^m \to \mathbb{R}_+$.

Fisher showed that if all the utility functions $u_i$ satisfy some mild conditions, then there always exists an equilibrium price vector $\mathbf{p} \in \mathbb{R}_+^m$. At this price, one can find a bundle of goods $\mathbf{a}_i$ for each buyer $i$ such that $\mathbf{a}_i$ maximizes her utility under the budget constraint that

$$\sum_{j \in [m]} a_{i,j} \cdot p_j \leq w_i,$$

and at the same time, the market demands equal to the market supply:

$$\sum_{i \in [n]} a_{i,j} \leq c_j, \quad \text{for all } j \in [m].$$

---

[3] Recently, Vazirani and Yannakakis independently proved that the problem of computing a Fisher equilibrium in a market with additively separable and PLC utility functions is PPAD-complete [1].

Fisher's model is a special case of the more general model of exchange economies considered by Arrow and Debreu [3]: In an exchange economy, there are $n$ traders and $m$ divisible goods. Trader $i$ has an *initial endowment* of $w_{i,j} \geq 0$ of good $j$ and a utility function $u_i : \mathbb{R}_+^m \to \mathbb{R}_+$. The individual goal of a trader is to obtain a new bundle of goods that maximizes her utility.

In a sense, Fisher's model focuses more on spending than trading as in Arrow-Debreu's model. In his model, money can be viewed as a special kind of good. All but one "special" trader only have money as their endowments, and money has no value to their utilities; the special trader, sometime called the "*market*", has all the goods and her interest is to collect all the money.

Over the last two decades, we have more algorithmic success in computing a market equilibrium in Fisher's model than computing an equilibrium in Arrow-Debreu's model.

– For the latter, polynomial-time algorithms are only known for markets with utility functions that are linear [4–12] or satisfy weak gross substitutability [13]. These algorithms critically used the fact that the set of equilibria of these markets is convex. Progress on markets with non-convex set of equilibria has been relatively slow. There are only a few algorithms in this case. Devanur and Kannan [14] gave a polynomial-time algorithm for markets with piecewise-linear and concave (PLC) utilities and a constant number of goods. Codenotti, McCune, Penumatcha, and Varadarajan [15] gave a polynomial-time algorithm for CES markets when the elasticity of substitution $s \geq 1/2$.

  For Leontief markets, in which each utility function is of the form $\min_j a_j x_j$, computing an approximate Arrow-Debreu equilibrium price is known to be PPAD-hard [16–18]. Recently, Chen *et. al.* [19] showed that finding an approximate equilibrium in an Arrow-Debreu exchange market, even if all the utility functions are additively separable [4] PLC, is complete in PPAD.

– For Fisher's model, polynomial-time algorithms are given not only for linear markets but also for Leontief and many other markets, e.g., the hybrid linear-Leontief markets [20]. We know that an (approximate) market equilibrium in any Fisher's economy with CES utilities can be found in polynomial time [4, 15, 12, 21, 7, 22]. In fact, Ye [21] proved that if every utility function is the minimum of a collection of homogeneous linear functions, then one can find a Fisher equilibrium in polynomial time.

### 1.2 Our Results

It remains open whether there is a family of concave utility functions for which it is PPAD-hard to compute a Fisher equilibrium. The family of utility functions that has drawn most attention is the additively separable, piecewise-linear, and concave (PLC) functions. Vazirani [23] remarked that obtaining a polynomial-time algorithm for markets with additively separable and concave utility func-tions is a premier open question today. Although the recent result of Chen *et. al.* [19] settled the complexity of computing an Arrow-Debreu equilibrium in markets with additively separable and PLC utilities, the complexity of Fisher equilibria remains unsettled.

---
[4] A function $u(x_1, \ldots, x_m)$ from $\mathbb{R}_+^m$ to $\mathbb{R}_+$ is *additively separable* if there exist $m$ real-valued functions $f_1, \ldots, f_m$ such that $u(x_1, \ldots, x_m) = \sum_{j=1}^m f_j(x_j)$.

In this paper, we show that the problem of finding a Fisher equilibrium remains to be PPAD-complete when the utility functions are additively separable and PLC. Therefore, for this seemingly simple class of utility functions, finding a Fisher equilibrium is as hard as finding an Arrow-Debreu equilibrium. Recently, Vazirani and Yannakakis independently proved that the problem of finding a Fisher equilibrium in a market with additively separable and PLC utility functions is PPAD-complete [1].

### 1.3 Sketch of the Proof

We prove the PPAD-hardness of computing a Fisher equilibrium by giving a reduction from SPARSE BIMATRIX [24]: the problem of computing an approximate Nash equilibrium in a sparse two-player game (see Section 2.1 for definition).

Similar to [19], our reduction starts by constructing a family of markets $\mathcal{M}_n$ for every $n \geq 1$, which we refer to as the *price-regulating* markets. There are $2n$ goods in $\mathcal{M}_n$, and every approximate equilibrium price $\mathbf{p}$ satisfies the following *price-regulation* property:

$$p_{2k-1} + p_{2k} \approx 3 \quad \text{and} \quad 1/2 \leq p_{2k-1}/p_{2k} \leq 2, \quad \text{for every } k \in [n].$$

This allows us to encode $n$ $[0,1]$-variables $x_1, \ldots, x_n$ using $\mathbf{p}$ as

$$x_k = p_{2k} - (p_{2k} + p_{2k+1})/3, \quad \text{for every } k \in [n]. \tag{1}$$

Moreover, the price-regulation property is *stable* with respect to "small perturbations" to $\mathcal{M}_n$: When new buyers are added to $\mathcal{M}_n$ (without introducing new goods), this property remains to hold as long as the total amount of money of these new buyers is small compared to that of the buyers in $\mathcal{M}_n$. We remark that the price-regulating markets $\mathcal{M}_n$ in this paper are different from those in [19], simply because we are dealing with Fisher's model. In particular, our family $\{\mathcal{M}_n\}$ is piecewise-linear while the one in [19] is linear.

Given an $n \times n$ two-player game $(\mathbf{A}, \mathbf{B})$, we construct a market $\mathcal{M}$ by adding new buyers to $\mathcal{M}_{2n+1}$ (with $4n+2$ goods). All the new buyers have very little money compared to those in $\mathcal{M}_{2n+1}$ so the price-regulation property still holds. This enables us to encode a pair of probability distributions $(\mathbf{x}, \mathbf{y})$ of dimension $n$ (with $2n$ variables) using the first $4n$ entries of the price vector $\mathbf{p}$ as in (1). By using the price-regulation property we show how to set the utility functions of the new buyers appropriately so that we can control their preferences over the goods and ultimately implement all the Nash equilibrium constraints over $(\mathbf{x}, \mathbf{y})$ through $\mathbf{p}$. As a result, given any (approximate) market equilibrium price $\mathbf{p}$ of $\mathcal{M}$, the pair $(\mathbf{x}, \mathbf{y})$ obtained (after normalization) must be an approximate Nash equilibrium of $(\mathbf{A}, \mathbf{B})$.

## 2 Preliminaries

### 2.1 Complexity of Nash Equilibria

A two-player game is defined by the payoff matrices $(\mathbf{A}, \mathbf{B})$ of its two players. In this paper, we assume that both players have $n$ choices of actions and thus, both $\mathbf{A}$ and $\mathbf{B}$ are square matrices with $n$ rows and columns. We use $\Delta^n \subset \mathbb{R}^n$ to denote the set of probability distributions of $n$ dimensions.

We say a pair of probability vectors $(\mathbf{x}, \mathbf{y})$, where $\mathbf{x} \in \Delta^n$ and $\mathbf{y} \in \Delta^n$, is a Nash equilibrium of $(\mathbf{A}, \mathbf{B})$ if for all $i$ and $j$ in $[n] = \{1, 2, \ldots, n\}$,

$$\mathbf{A}_i \mathbf{y}^T > \mathbf{A}_j \mathbf{y}^T \implies x_j = 0 \quad \text{and} \quad \mathbf{x}\mathbf{B}_i > \mathbf{x}\mathbf{B}_j \implies y_j = 0,$$

where we use $\mathbf{A}_i$ to denote the $i$th row vector of $\mathbf{A}$, and $\mathbf{B}_i$ to denote the $i$th column vector of $\mathbf{B}$, respectively. For $\epsilon > 0$, $(\mathbf{x}, \mathbf{y})$ is an $\epsilon$-well-supported Nash equilibrium of $(\mathbf{A}, \mathbf{B})$, if $\mathbf{x}, \mathbf{y} \in \Delta^n$ and for all $i, j \in [n]$,

$$\mathbf{A}_i \mathbf{y}^T - \mathbf{A}_j \mathbf{y}^T > \epsilon \implies x_j = 0 \quad \text{and} \quad \mathbf{x}\mathbf{B}_i - \mathbf{x}\mathbf{B}_j > \epsilon \implies y_j = 0. \qquad (2)$$

A two-player game $(\mathbf{A}, \mathbf{B})$ is said to be *normalized* if every entry of $\mathbf{A}$ and $\mathbf{B}$ is between $-1$ and $1$. We say a two-player game $(\mathbf{A}, \mathbf{B})$ is *sparse* if every row and every column of $\mathbf{A}$ and $\mathbf{B}$ have at most 10 nonzero entries.

Let SPARSE BIMATRIX denote the following search problem: given an $n \times n$ sparse normalized two-player game, find an $n^{-6}$-well-supported Nash equilibrium. By [24], we know that SPARSE BIMATRIX is PPAD-complete.

### 2.2 Markets with Additively Separable PLC Utilities

Let $\mathcal{G} = \{G_1, \ldots, G_n\}$ denote a set of $n$ divisible goods, and $\mathcal{T} = \{T_1, \ldots, T_m\}$ denote a set of buyers. For each good $G_j$, we use $c_j > 0$ to denote the amount of $G_j$ in the market. For each buyer $T_i$, we use $w_i > 0$ to denote her money and $u_i : \mathbb{R}_+^n \to \mathbb{R}_+$ to denote her utility function. In this paper, we will mainly focus on markets with additively separable, piecewise-linear and concave utilities.

A continuous function $r(\cdot)$ over $\mathbb{R}_+$ is said to be *$t$-segment* piecewise linear and concave (PLC) if $r(0) = 0$ and there exists a tuple

$$[\theta_0 > \theta_1 > \ldots > \theta_t \geq 0; 0 < a_1 < a_2 < \ldots < a_t]$$

of length $2t + 1$ such that (letting $a_0 = 0$)

1. $\forall i \in [0 : t-1]$, the restriction of $r(\cdot)$ over $[a_i, a_{i+1}]$ is a segment of slope $\theta_i$;
2. the restriction of $r(\cdot)$ over $[a_t, +\infty)$ is a ray of slope $\theta_t$.

The $(2t+1)$-tuple is called the *representation* of $r(\cdot)$. Also we say $r(\cdot)$ is *strictly monotone* if $\theta_t > 0$, and is *$\alpha$-bounded* for some $\alpha \geq 1$ if $\alpha \geq \theta_0$ and $\theta_t \geq 1$.

**Definition 1.** *A function $u(\cdot) : \mathbb{R}_+^n \to \mathbb{R}_+$ is said to be an* additively separable PLC *function if there exist PLC functions $r_1(\cdot), \ldots, r_n(\cdot) : \mathbb{R}_+ \to \mathbb{R}_+$ such that*

$$u(\mathbf{a}) = \sum_{j \in [n]} r_j(a_j), \quad \text{for all } \mathbf{a} \in \mathbb{R}_+^n.$$

In such a market, we use, for each buyer $T_i \in \mathcal{T}$, $r_{i,j}(\cdot) : \mathbb{R}_+ \to \mathbb{R}_+$ to denote her PLC function with respect to good $G_j \in \mathcal{G}$. As a result, we have

$$u_i(\mathbf{a}) = \sum_{j \in [n]} r_{i,j}(a_j), \quad \text{for all } \mathbf{a} \in \mathbb{R}_+^n.$$

We use $\mathbf{p} \in \mathbb{R}_+^n$ to denote a price vector, where $\mathbf{p} \neq \mathbf{0}$ and $p_j$ is the price of $G_j$. Given $\mathbf{p}$, we let $\text{OPT}(i, \mathbf{p})$ denote the set of allocations that maximize $u_i(\cdot)$:

$$\text{OPT}(i, \mathbf{p}) = \text{argmax}_{\mathbf{a} \in \mathbb{R}_+^n,\ \mathbf{a} \cdot \mathbf{p} \leq w_i}\ u_i(\mathbf{a}).$$

We let $\mathcal{X} = \{\mathbf{a}_i \in \mathbb{R}_+^n : i \in [m]\}$ denote an allocation of the market: for each buyer $T_i \in \mathcal{T}$, $\mathbf{a}_i \in \mathbb{R}_+^n$ is the amount of goods that $T_i$ receives. In particular, the amount of $G_j$ that $T_i$ receives in $\mathcal{X}$ is $a_{i,j}$.

**Definition 2.** *A market equilibrium is a nonzero vector $\mathbf{p} \in \mathbb{R}_+^n$ such that there exists an allocation $\mathcal{X} = \{\mathbf{a}_i : i \in [m]\}$ which has the following two properties:*

1. *Every buyer gets an optimal bundle: for every $T_i$, $\mathbf{a}_i \in \text{OPT}(i, \mathbf{p})$;*
2. *The market clears: for every $G_j \in \mathcal{G}$, $\sum_{i \in [m]} a_{i,j} \leq c_j$. In particular,*

$$p_j > 0 \implies \sum_{i \in [m]} a_{i,j} = c_j.$$

In general, not every market $\mathcal{M}$ has such an equilibrium price vector. However, for the additively separable PLC markets, the following condition guarantees the existence of an equilibrium:

> If for every buyer $T_i \in \mathcal{T}$ there exists a good $G_j \in \mathcal{G}$ such that the PLC function $r_{i,j}(\cdot)$ is *strictly monotone*, then a market equilibrium $\mathbf{p}$ exists.

It is a corollary of Maxfield [25]. Moreover, one can show that (e.g., see [14, 19]) if all the parameters of $\mathcal{M}$ are rational numbers, then it must have a rational equilibrium $\mathbf{p}$, and the number of bits needed to describe $\mathbf{p}$ is polynomial in the input size of $\mathcal{M}$ (i.e., the number of bits we need to describe the market $\mathcal{M}$).

We are interested in the problem of finding an approximate market equilibrium in an additively separable PLC market.

**Definition 3 (Approximate Market Equilibrium).** *Let $\mathcal{M}$ be an additively separable PLC market. We say $\mathbf{p}$ is an $\epsilon$-approximate market equilibrium of $\mathcal{M}$, for some $\epsilon \geq 0$, if there is an allocation $\mathcal{X} = \{\mathbf{a}_i \in \mathbb{R}_+^n : i \in [m]\}$ such that every buyer gets an optimal bundle with respect to $\mathbf{p}$: $\mathbf{a}_i \in \text{OPT}(i, \mathbf{p})$, for all $i \in [m]$; and the market clears approximately: for all $G_j \in \mathcal{G}$,*

$$\left| \sum_{i \in [m]} a_{i,j} - c_j \right| \leq \epsilon \cdot c_j.$$

We make some further restrictions on the markets we are interested in. We say an additively separable PLC market $\mathcal{M}$ is $\alpha$-bounded, for some $\alpha \geq 1$, if for all $T_i$ and $G_j$, the PLC function $r_{i,j}(\cdot)$ is either the zero function or $\alpha$-bounded. We call an additively separable PLC market $\mathcal{M}$ a 2-linear market, if for all $T_i$ and $G_j$, $r_{i,j}(\cdot)$ has at most two segments. Finally we say an additively separable PLC market $\mathcal{M}$ is $t$-sparse, for some positive integer $t$, if for any $T_i$, the number of $j \in [n]$ such that $r_{i,j}(\cdot)$ is not the zero function is at most $t$. In another word, every buyer $T_i$ is interested in at most $t$ goods.

We use FISHER to denote the following search problem: given a 2-linear additively separable PLC market $\mathcal{M}$, which is 81-bounded, 43-sparse and satisfies the condition of Maxfield, find an $n^{-21}$-approximate market equilibrium, where $n$ denotes the number of goods in the market. It is not hard to show that FISHER is in PPAD (e.g., see [19]). The main result of the paper is to show that FISHER is actually PPAD-complete.

**Theorem 1 (Main).** FISHER *is* PPAD-complete.

## 3 A Price-Regulating Market

In this section, we construct a family of *price-regulating* markets $\{\mathcal{M}_n : n \geq 1\}$ in Fisher's setting. For every positive integer $n$, $\mathcal{M}_n$ has $n$ buyers, $2n$ goods and satisfies the following price regulation property.

*Property 1 (**Price Regulation**). Let $\mathbf{p} \in \mathbb{R}_+^{2n}$ be an $\epsilon$-approximate equilibrium of $\mathcal{M}_n$ with $\epsilon < 1$, then we have*

$$\frac{3}{1+\epsilon} \leq p_{2k-1} + p_{2k} \leq \frac{3}{1-\epsilon} \quad \text{and} \quad \frac{1}{2} \leq \frac{p_{2k-1}}{p_{2k}} \leq 2, \quad \text{for every } k \in [n].$$

We start with some notation. The goods in $\mathcal{M}_n$ are $\mathcal{G} = \{G_1, \ldots, G_{2n}\}$ and the buyers in $\mathcal{M}_n$ are $\mathcal{T} = \{T_1, \ldots, T_n\}$. For each buyer $T_i \in \mathcal{T}$, we use $w_i > 0$ to denote her money, $u_i(\cdot)$ to denote her utility function, $r_{i,k}(\cdot)$ to denote her PLC function with respect to $G_k$, and $\text{OPT}(i, \mathbf{p})$ to denote the set of bundles that maximize her utility with respect to $\mathbf{p}$.

In the construction of $\mathcal{M}_n$ below, we use $r(\cdot) \Leftarrow [\theta]$ to denote the action of setting $r(\cdot)$ to be the linear function of slope $\theta \geq 0$; and use $r(\cdot) \Leftarrow [\theta_0, \theta_1; a_1]$ to denote the action of setting it to be the 2-segment function with representation $[\theta_0, \theta_1; a_1]$, where $\theta_0 > \theta_1$ and $a_1 > 0$.

**Construction of $\mathcal{M}_n$:** First, we set $c_k = 1$ for all $k \in [2n]$. Second, for every $i \in [n]$, we set $w_i = 3$. Finally, we set the PLC functions $r_{i,k}(\cdot)$ as follows:

1. For all $k \neq 2i-1, 2i$, we set $r_{i,k}(\cdot)$ to be the zero function: $r_{i,k}(\cdot) \Leftarrow [0]$;
2. $r_{i,2i-1}(\cdot) \Leftarrow [2]$; and $r_{i,2i}(\cdot) \Leftarrow [4, 1; 1]$.

This finishes the construction of $\mathcal{M}_n$ (which is 2-linear, 4-bounded and 2-sparse).

*Proof (Proof of Property 1).* Let $\mathbf{p}$ be an $\epsilon$-approximate equilibrium, and $\mathcal{X} = \{\mathbf{a}_i \in \mathbb{R}_+^{2n} : i \in [n]\}$ be an optimal allocation that clears the market approximately. Without loss of generality, we prove Property 1 for $k = 1$.

First, it is easy to check that $p_1, p_2 > 0$ since otherwise, we have $a_{1,1} = +\infty$ or $a_{1,2} = +\infty$, which contradicts the assumption of $\mathbf{p}$ being an $\epsilon$-approximate market equilibrium.

Second, we show that $p_1/p_2 \le 2$. Assume, for contradiction, that $p_1 > 2 \cdot p_2$. By the optimality of $\mathbf{a}_1$, we have $a_{1,1} = 0$. As a result, we have $a_{i,1} = 0$ for all $i \in [n]$, which contradicts the assumption that $\mathbf{p}$ is an approximate equilibrium. Similarly, one can show that $p_1/p_2 \ge 1/2$.

Finally, by the optimality of $\mathbf{a}_1$, we have $3 = a_{1,1} \cdot p_1 + a_{1,2} \cdot p_2$. Since $\mathbf{p}$ is an $\epsilon$-approximate market equilibrium, we have $|a_{1,1} - 1|, |a_{1,2} - 1| \le \epsilon$. As a result,

$$(1 - \epsilon)(p_1 + p_2) \le 3 = a_{1,1} \cdot p_1 + a_{1,2} \cdot p_2 \le (1 + \epsilon)(p_1 + p_2)$$

and Lemma 1 follows.

By Lemma 1, we have

$$p_{2k-1}, p_{2k} \in \left[\frac{p_{2k-1} + p_{2k}}{3}, \frac{2(p_{2k-1} + p_{2k})}{3}\right] \subset \left[\frac{1}{1+\epsilon}, \frac{2}{1-\epsilon}\right].$$

In the next section, we use $\mathcal{M}_{2n+1}$ and the following $2n$ variables derived from $\mathbf{p}$ to encode a pair of $n$-dimensional distributions $(\mathbf{x}, \mathbf{y})$: For $k \in [n]$,

$$x_k = p_{2k} - (p_{2k-1} + p_{2k})/3 \quad \text{and} \quad y_k = p_{2(n+k)} - (p_{2(n+k)-1} + p_{2(n+k)})/3.$$

Given an $n \times n$ sparse two-player game $(\mathbf{A}, \mathbf{B})$, we show how to add new buyers to "perturb" the market $\mathcal{M}_{2n+1}$ so that any approximate equilibrium $\mathbf{p}$ of the new market yields an approximate Nash equilibrium $(\mathbf{x}, \mathbf{y})$ of $(\mathbf{A}, \mathbf{B})$.

## 4 Reduction from SPARSE BIMATRIX to FISHER

In this section we prove Theorem 1 by giving a polynomial-time reduction from SPARSE BIMATRIX to FISHER. Given an $n \times n$ sparse two-player game $(\mathbf{A}, \mathbf{B})$, where $\mathbf{A}, \mathbf{B} \in [-1, 1]^{n \times n}$, we build an additively separable PLC market $\mathcal{M}$ by adding more buyers to the price-regulating market $\mathcal{M}_{2n+1}$. There are $4n + 2$ goods $\mathcal{G} = \{G_1, \ldots, G_{4n}, G_{4n+1}, G_{4n+2}\}$ in $\mathcal{M}$, and the buyers $\mathcal{T}$ in $\mathcal{M}$ are

$$\mathcal{T} = \{T_i, T_\mathbf{u}, T_\mathbf{v} : i \in [2n + 1], \mathbf{u} \in U \text{ and } \mathbf{v} \in V\},$$

where $U = \{(i, j, 1) : 1 \le i \ne j \le n\}$ and $V = \{(i, j, 2) : 1 \le i \ne j \le n\}$. The buyers $\{T_i\}$ have almost the same money and PLC functions as in $\mathcal{M}_{2n+1}$.

When constructing the market $\mathcal{M}$, we also define a $4n$-dimensional vector $\mathbf{s}_\mathbf{u}$ for every buyer $T_\mathbf{u}$, and a $4n$-dimensional vector $\mathbf{s}_\mathbf{v}$ for every buyer $T_\mathbf{v}$, which will be useful in the proof of correctness.

### 4.1 Setting up the Market $\mathcal{M}$

First, we set the money and utility function of each buyer $T \in \mathcal{T}$.

**Buyers $T_i$, where $i \in [2n+1]$.** For every $T_i \in \mathcal{T}$, where $i \in [2n+1]$, we set her money $w_i$ and PLC functions $r_{i,k}(\cdot)$ almost the same as in $\mathcal{M}_{2n+1}$. First we set $w_i = 3$. Second, the PLC function $r_{i,k}(\cdot)$ is set as:

1. $r_{i,k}(\cdot) \Leftarrow [0]$ for all $k \neq 2i-1, 2i$; and
2. $r_{i,2i-1}(\cdot) \Leftarrow [2]$; and $r_{i,2i}(\cdot) \Leftarrow [4, 1; 1 + 1/n^{20}]$.

**Buyers $T_{\mathbf{u}}$, where $\mathbf{u} \in U$.** Let $\mathbf{u} = (i, j, 1)$, where $1 \leq i \neq j \leq n$. We use $\mathbf{A}_i$ and $\mathbf{A}_j$ to denote the $i$th and $j$th row vectors of $\mathbf{A}$, respectively, and use $\mathbf{C}$ to denote $\mathbf{A}_i - \mathbf{A}_j$. Because $\mathbf{A} \in [-1,1]^{n \times n}$, $|C_k| \leq 2$ for all $k$. We denote by $m$ the number of nonzero entries in $\mathbf{C}$, then it is clear that $m \leq 20$. Let $C = \sum_{k \in [n]} C_k$ then we have $|C| \leq 20$.

First, we set the money $w_{\mathbf{u}}$ of $T_{\mathbf{u}}$ to be

$$w_{\mathbf{u}} = \frac{3}{n^{12}} + \frac{6m + C}{n^{13}}.$$

Using $\mathbf{C}$, we set the PLC functions $r_{\mathbf{u},k}(\cdot)$, where $k \in [4n+2]$, of $T_{\mathbf{u}}$ as follows:

1. $r_{\mathbf{u},2(n+k)-1}(\cdot) \Leftarrow [0]$ and $r_{\mathbf{u},2(n+k)}(\cdot) \Leftarrow [0]$ for all $k \in [n]$ such that $C_k = 0$;
2. $r_{\mathbf{u},2(n+k)-1}(\cdot) \Leftarrow [81, 1; 2/n^{13}]$ for all $k \in [n]$ such that $C_k \neq 0$;
3. $r_{\mathbf{u},2(n+k)}(\cdot) \Leftarrow [81, 1; (2 + C_k)/n^{13}]$ for all $k \in [n]$ such that $C_k \neq 0$;
4. $r_{\mathbf{u},2j-1}(\cdot) \Leftarrow [27, 1; 1/n^{12}]$ and $r_{\mathbf{u},2j}(\cdot) \Leftarrow [9, 1; 1/n^{12}]$;
5. $r_{\mathbf{u},k}(\cdot) \Leftarrow [0]$ for all other $k \in [2n]$;
6. $r_{\mathbf{u},4n+1}(\cdot) \Leftarrow [3]$ and $r_{\mathbf{u},4n+2}(\cdot) \Leftarrow [0]$.

We also define the auxiliary vector $\mathbf{s}_{\mathbf{u}} \in \mathbb{R}_+^{4n}$ as follows:

1. $s_{\mathbf{u},2(n+k)-1} = s_{\mathbf{u},2(n+k)} = 0$ for all $k \in [n]$ such that $C_k = 0$;
2. $s_{\mathbf{u},2(n+k)-1} = 2/n^{13}$ and $s_{\mathbf{u},2(n+k)} = (2 + C_k)/n^{13}$ for all $k$ with $C_k \neq 0$;
3. $s_{\mathbf{u},2j-1} = s_{\mathbf{u},2j} = 1/n^{12}$; and
4. $s_{\mathbf{u},k} = 0$ for all other $k \in [2n]$.

**Buyers $T_{\mathbf{v}}$, where $\mathbf{v} \in V$.** The behavior of $T_{\mathbf{v}}$, $\mathbf{v} \in V$, is similar to that of $T_{\mathbf{u}}$ except that it works on the second payoff matrix $\mathbf{B}$.

Let $\mathbf{v} = (i, j, 2) \in V$, where $1 \leq i \neq j \leq n$. We let $\mathbf{B}_i$ and $\mathbf{B}_j$ denote the $i$th and $j$th column vectors of $\mathbf{B}$, respectively, and use $\mathbf{C}$ to denote $\mathbf{B}_i - \mathbf{B}_j$. We also use $m$ to denote the number of nonzero entries in $\mathbf{C}$ and $C$ to denote $\sum_{k \in [n]} C_k$.

First, we set the money $w_{\mathbf{v}} > 0$ of $T_{\mathbf{v}}$ to be

$$w_{\mathbf{v}} = \frac{3}{n^{12}} + \frac{6m + C}{n^{13}}.$$

Using **C**, we set the utility functions $r_{\mathbf{v},k}(\cdot)$, where $k \in [4n+2]$, of $T_{\mathbf{v}}$ as follows:

1. $r_{\mathbf{v},2k-1}(\cdot) \Leftarrow [0]$ and $r_{\mathbf{v},2k}(\cdot) \Leftarrow [0]$ for all $k \in [n]$ such that $C_k = 0$;
2. $r_{\mathbf{v},2k-1}(\cdot) \Leftarrow [81, 1; 2/n^{13}]$ for all $k \in [n]$ such that $C_k \neq 0$;
3. $r_{\mathbf{v},2k}(\cdot) \Leftarrow [81, 1; (2+C_k)/n^{13}]$ for all $k \in [n]$ such that $C_k \neq 0$;
4. $r_{\mathbf{v},2(n+j)-1}(\cdot) \Leftarrow [27, 1; 1/n^{12}]$ and $r_{\mathbf{v},2(n+j)}(\cdot) \Leftarrow [9, 1; 1/n^{12}]$;
5. $r_{\mathbf{v},k}(\cdot) \Leftarrow [0]$ for all other $k \in [2n:4n]$;
6. $r_{\mathbf{v},4n+1}(\cdot) \Leftarrow [3]$ and $r_{\mathbf{v},4n+2}(\cdot) \Leftarrow [0]$.

Similarly, we define the auxiliary vector $\mathbf{s_v} \in \mathbb{R}_+^{4n}$ as follows:

1. $s_{\mathbf{v},2k-1} = s_{\mathbf{v},2k} = 0$ for all $k \in [n]$ such that $C_k = 0$;
2. $s_{\mathbf{v},2k-1} = 2/n^{13}$ and $s_{\mathbf{v},2k} = (2+C_k)/n^{13}$ for all $k \in [n]$ such that $C_k \neq 0$;
3. $s_{\mathbf{v},2(n+j)-1} = s_{\mathbf{v},2(n+j)} = 1/n^{12}$; and
4. $s_{\mathbf{v},k} = 0$ for all other $k \in [2n:4n]$.

**Setting $c_k$, where $k \in [4n+2]$.** First, $c_{4n+1} = c_{4n+2} = 1$. Second, we set

$$c_k = 1 + \sum_{\mathbf{u} \in U} s_{\mathbf{u},k} + \sum_{\mathbf{v} \in V} s_{\mathbf{v},k}, \quad \text{for every } k \in [4n],$$

using the auxiliary vectors $\mathbf{s_u}$ and $\mathbf{s_v}$. This finishes the construction of $\mathcal{M}$. It is easy to check that the market $\mathcal{M}$ constructed is 2-linear, 81-bounded, 43-sparse, and satisfies the condition of Maxfield.

### 4.2 Sketch of the Reduction

Let $N = 4n+2$, the number of goods in $\mathcal{M}$. Then to prove Theorem 1, we only need to show that from every $N^{-21}$-approximate equilibrium $\mathbf{p}$ of $\mathcal{M}$, one can construct an $n^{-6}$-well-supported equilibrium $(\mathbf{x}, \mathbf{y})$ of $(\mathbf{A}, \mathbf{B})$ efficiently.

To this end, we let $(\mathbf{x}', \mathbf{y}')$ denote the following two $n$-dimensional vectors:

$$x'_k = p_{2k} - \frac{p_{2k-1} + p_{2k}}{3} \quad \text{and} \quad y'_k = p_{2(n+k)} - \frac{p_{2(n+k)-1} + p_{2(n+k)}}{3}. \tag{3}$$

Then we normalize $(\mathbf{x}', \mathbf{y}')$ to get $(\mathbf{x}, \mathbf{y})$ (we will show later that $\mathbf{x}', \mathbf{y}' \neq \mathbf{0}$):

$$x_k = \frac{x'_k}{\sum_{i \in [n]} x'_i} \quad \text{and} \quad y_k = \frac{y'_k}{\sum_{i \in [n]} y'_i}, \quad \text{for every } k \in [n]. \tag{4}$$

Theorem 1 then follows from Theorem 2 below, which we will prove in the next section. Note that if $\mathbf{p}$ is an $N^{-21}$-approximate equilibrium, then by definition it is also an $n^{-21}$-approximate equilibrium.

**Theorem 2.** *If $\mathbf{p}$ is an $n^{-21}$-approximate market equilibrium of $\mathcal{M}$, then $(\mathbf{x}, \mathbf{y})$ constructed above must be an $n^{-6}$-well-supported Nash equilibrium of $(\mathbf{A}, \mathbf{B})$.*

## 5 Correctness of the Reduction

In this section, we prove Theorem 2. Let $\mathbf{p} = (p_1, \ldots, p_{4n+2})$ be an $n^{-21}$-approximate equilibrium of $\mathcal{M}$. It is easy to show that $p_k > 0$ for all $k$. Let $\mathcal{X}$ be an optimal allocation with respect to $\mathbf{p}$ that clears the market approximately:

$$\mathcal{X} = \big\{\mathbf{a}_i, \mathbf{a_u}, \mathbf{a_v} \in \mathbb{R}_+^{4n+2} : i \in [2n+1], \mathbf{u} \in U \text{ and } \mathbf{v} \in V\big\}.$$

We start with some notation. We let

$$\mathcal{T}^* = \big\{T_i : i \in [2n+1]\big\}, \quad \mathcal{T}_U = \big\{T_\mathbf{u} : \mathbf{u} \in U\big\} \quad \text{and} \quad \mathcal{T}_V = \big\{T_\mathbf{v} : \mathbf{v} \in V\big\}.$$

Let $\mathcal{T}' \subseteq \mathcal{T}$ be a subset of buyers and $k \in [4n+2]$, then we use $a_k[\mathcal{T}']$ to denote the amount of good $G_k$ that buyers in $\mathcal{T}'$ receive in the final allocation $\mathcal{X}$. For $\mathcal{T}' \subseteq \mathcal{T}_U \cup \mathcal{T}_V$ and $k \in [4n]$, we let

$$s_k[\mathcal{T}'] = \sum_{T_\mathbf{u} \in \mathcal{T}' \cap \mathcal{T}_U} s_{\mathbf{u},k} + \sum_{T_\mathbf{v} \in \mathcal{T}' \cap \mathcal{T}_V} s_{\mathbf{v},k}.$$

By the construction of $\mathcal{M}$, we have $c_{4n+1} = c_{4n+2} = 1$ and

$$1 < c_k = 1 + \Theta(1/n^{11}) < 2, \quad \text{for every } k \in [4n].$$

By the definition of approximate equilibria, $|c_k - a_k[\mathcal{T}]| \le c_k/n^{21} < 2/n^{21}$ and

$$\big|s_k[\mathcal{T}_\mathbf{u} \cup \mathcal{T}_\mathbf{v}] - a_k[\mathcal{T}_\mathbf{u} \cup \mathcal{T}_\mathbf{v}] + 1 - a_k[\mathcal{T}^*]\big| < 2/n^{21}, \quad \text{for all } k \in [4n]. \quad (5)$$

### 5.1 The Price-Regulation Property

First we show that the price vector $\mathbf{p}$ must satisfy the following price-regulation property. The proof is similar to that of Property 1, which mainly uses the fact that buyers in $\mathcal{T}^*$ possess almost all the money in the market $\mathcal{M}$.

**Lemma 1 (Price Regulation).** *For every $k \in [2n+1]$, we have*

$$\frac{1}{2} \le \frac{p_{2k-1}}{p_{2k}} \le 2 \quad \text{and} \quad 3 - O\left(\frac{1}{n^{11}}\right) \le p_{2k-1} + p_{2k} \le 3 + O\left(\frac{1}{n^{10}}\right).$$

*Proof.* We start with the second part of the lemma.

First, the total money that buyers in $\mathcal{T}$ spend on $G_{2k-1}$ and $G_{2k}$ is

$$p_{2k-1} \cdot a_{2k-1}[\mathcal{T}] + p_{2k} \cdot a_{2k}[\mathcal{T}] \le 3 + O(1/n^{12}) \cdot (|U| + |V|) = 3 + O(1/n^{10}) \quad (6)$$

since buyers $T_i$, $i \ne k$, are not interested in $G_{2k-1}$ and $G_{2k}$. On the other hand, because $\mathbf{p}$ is an approximate equilibrium, we have

$$a_{2k-1}[\mathcal{T}] \ge c_{2k-1} \cdot (1 - 1/n^{21}) \ge 1 - 1/n^{21} \quad \text{and} \quad a_{2k}[\mathcal{T}] \ge 1 - 1/n^{21}.$$

As a result,

$$p_{2k-1} \cdot a_{2k-1}[\mathcal{T}] + p_{2k} \cdot a_{2k}[\mathcal{T}] \ge (p_{2k-1} + p_{2k})(1 - 1/n^{21}). \quad (7)$$

By combining (6) and (7), we have $p_{2k-1} + p_{2k} \leq 3 + O(1/n^{10})$.

Second, by the optimality of $\mathbf{a}_k$, we have

$$3 = p_{2k-1} \cdot a_{k,2k-1} + p_{2k} \cdot a_{k,2k}. \tag{8}$$

On the other hand, because $\mathbf{p}$ is an approximate equilibrium, we have

$$a_{k,2k-1} \leq a_{2k-1}[\mathcal{T}] \leq c_{2k-1}(1 + 1/n^{21}) = 1 + O(1/n^{11})$$

and $a_{k,2k} \leq 1 + O(1/n^{11})$. As a result,

$$p_{2k-1} \cdot a_{k,2k-1} + p_{2k} \cdot a_{k,2k} \leq (p_{2k-1} + p_{2k})\big(1 + O(1/n^{11})\big). \tag{9}$$

By combining (8) and (9), we have $p_{2k-1} + p_{2k} \geq 3 - O(1/n^{11})$.

Finally we prove the first part of the lemma. Assume, for contradiction, that $p_{2k-1} > 2 \cdot p_{2k}$ for some $k \in [2n+1]$. By the optimality of $\mathbf{a}_k$, $a_{k,2k-1} = 0$ and thus, the money that buyers in $\mathcal{T}$ spend on $G_{2k-1}$ is at most

$$O(1/n^{12}) \cdot (|U| + |V|) = O(1/n^{10}).$$

However, since $p_{2k-1} > 2 \cdot p_{2k}$, the price of $G_{2k-1}$ is at least

$$2(p_{2k-1} + p_{2k})/3 \geq 2 - O(1/n^{11}),$$

which contradicts the assumption that $\mathbf{p}$ is an approximate equilibrium of $\mathcal{M}$. Similarly, one can show that $p_{2k} \leq 2 \cdot p_{2k-1}$, and the lemma is proven.

**Corollary 1.** *For all $i, j \in [4n+2]$, we have $p_i/p_j < 3$.*

Using Corollary 1, we analyze the behavior of $T_{\mathbf{u}}$ and $T_{\mathbf{v}}$ as follows.

**Behavior of $T_{\mathbf{u}}$:** Let $\mathbf{u} = (i, j, 1) \in U$, where $1 \leq i \neq j \leq n$. Let $\mathbf{C} = \mathbf{A}_i - \mathbf{A}_j$, $m \leq 20$ be the number of nonzero entries in $\mathbf{C}$ and $C = \sum_{k \in [n]} C_k$. By Corollary 1 and the optimality of $\mathbf{a}_{\mathbf{u}}$, $T_{\mathbf{u}}$ first buys the following bundle of goods:

$$\Big\{ s_{\mathbf{u},2(n+k)-1} \text{ of } G_{2(n+k)-1} \text{ and } s_{\mathbf{u},2(n+k)} \text{ of } G_{2(n+k)} : k \in [n] \text{ and } C_k \neq 0 \Big\}. \tag{10}$$

The money of $T_{\mathbf{u}}$ left is (we let $I$ denote the set of $k \in [n]$ such that $C_k \neq 0$)

$$\frac{3}{n^{12}} + \frac{6m + C}{n^{13}} - \frac{2}{n^{13}} \sum_{k \in I} \big(p_{2(n+k)-1} + p_{2(n+k)}\big) - \frac{1}{n^{13}} \sum_{k \in I} C_k \cdot p_{2(n+k)}. \tag{11}$$

By Lemma 1, the money left is $3/n^{12} - O(1/n^{13}) > 0$. After this, $T_{\mathbf{u}}$ buys $G_{2j-1}$ up to $1/n^{12}$ and the money left is $\Omega(1/n^{12})$ by Lemma 1. Finally, $T_{\mathbf{u}}$ buys $G_{2j}$ up to $1/n^{12}$ and spends all the money left, if any, to buy $G_{4n+1}$.

**Behavior of $T_{\mathbf{v}}$:** Let $\mathbf{v} = (i, j, 2) \in V$, where $1 \leq i \neq j \leq n$. Let $\mathbf{C} = \mathbf{B}_i - \mathbf{B}_j$, $m \leq 20$ be the number of nonzero entries in $\mathbf{C}$ and $C = \sum_{k \in [n]} C_k$. By Corollary 1 and the optimality of $\mathbf{a}_{\mathbf{v}}$, $T_{\mathbf{v}}$ first buys the following bundle of goods

$$\Big\{s_{\mathbf{v},2k-1} \text{ of } G_{2k-1} \text{ and } s_{\mathbf{v},2k} \text{ of } G_{2k} : k \in [n] \text{ with } C_k \neq 0\Big\}.$$

The money of $T_{\mathbf{v}}$ left is (we let $I$ denote the set of $k \in [n]$ such that $C_k \neq 0$)

$$\frac{3}{n^{12}} + \frac{6m + C}{n^{13}} - \frac{2}{n^{13}} \sum_{k \in I} (p_{2k-1} + p_{2k}) - \frac{1}{n^{13}} \sum_{k \in I} C_k \cdot p_{2k}. \tag{12}$$

By Lemma 1 the money left is $3/n^{12} - O(1/n^{13})$. After this, $T_{\mathbf{v}}$ buys $G_{2(n+j)-1}$ up to $1/n^{12}$ and the money left is $\Omega(1/n^{12})$. Finally, $T_{\mathbf{v}}$ buys good $G_{2(n+j)}$ up to $1/n^{12}$ and spends all the money left, if any, on $G_{4n+1}$.

The analysis above gives us the following corollary.

**Corollary 2.** *For all $k \in [2n]$, $a_{2k-1}[\mathcal{T}_U \cup \mathcal{T}_V] = s_{2k-1}[\mathcal{T}_U \cup \mathcal{T}_V]$. For all $T \in \mathcal{T}_U \cup \mathcal{T}_V$ and $k \in [2n]$, we have $a_{2k}[T] \leq s_{2k}[T]$. In particular,*

$$s_{2k}[\mathcal{T}_U \cup \mathcal{T}_V] - a_{2k}[\mathcal{T}_U \cup \mathcal{T}_V] \geq s_{2k}[T] - a_{2k}[T], \qquad \textit{for any } T \in \mathcal{T}_U \cup \mathcal{T}_V.$$

*For all $k \in [n]$, we have $s_{2k}[\mathcal{T}_V] = a_{2k}[\mathcal{T}_V]$ and $s_{2(n+k)}[\mathcal{T}_U] = a_{2(n+k)}[\mathcal{T}_U]$.*

By combining (5) and the first part of Corollary 2, we have,

$$\big|a_{k,2k-1} - 1\big| < 2/n^{21}, \qquad \text{for every } k \in [2n]. \tag{13}$$

### 5.2 Two Useful Lemmas

We prove two useful relations between $p_{2k}$ and $s_{2k}[\mathcal{T}_U \cup \mathcal{T}_V] - a_{2k}[\mathcal{T}_U \cup \mathcal{T}_V]$.

**Lemma 2.** *Let $\mathbf{p}$ be an $n^{-21}$-approximate market equilibrium of $\mathcal{M}$. If*

$$s_{2k}[\mathcal{T}_U \cup \mathcal{T}_V] - a_{2k}[\mathcal{T}_U \cup \mathcal{T}_V] = \Omega(1/n^{19}) \tag{14}$$

*for some $k \in [2n]$, then $p_{2k} = (p_{2k-1} + p_{2k})/3$.*

*Proof.* Assume (14) holds for some $k \in [2n]$. Then by (5) we have $a_{2k}[\mathcal{T}^*] - 1 = \Omega(1/n^{19})$ and thus, $a_{2k}[\mathcal{T}^*] > 1 + 1/n^{20}$. This implies that $a_{k,2k} > 1 + 1/n^{20}$. By (13) and the optimality of $\mathbf{a}_k$, we have $p_{2k-1}/2 = p_{2k}/1$.

**Lemma 3.** *Let $\mathbf{p}$ be an $n^{-21}$-approximate market equilibrium. If*

$$s_{2k}[\mathcal{T}_U \cup \mathcal{T}_V] - a_{2k}[\mathcal{T}_U \cup \mathcal{T}_V] = O(1/n^{21})$$

*for some $k \in [2n]$, then $p_{2k} = 2(p_{2k-1} + p_{2k})/3$.*

*Proof.* Assume (14) holds for some $k \in [2n]$. Then by (5) we have $a_{2k}[\mathcal{T}^*] - 1 \leq O(1/n^{21})$ and thus, $a_{2k}[\mathcal{T}^*] < 1 + 1/n^{20}$. This implies that $a_{k,2k} < 1 + 1/n^{20}$. By (13) and the optimality of $\mathbf{a}_k$, we have $p_{2k-1}/2 = p_{2k}/4$.

### 5.3 Proof of Theorem 2

Let $\mathbf{x}'$ and $\mathbf{y}'$ denote the two vectors obtained from $\mathbf{p}$ as in (3). By Lemma 1,

$$0 \leq x'_k, y'_k \leq 1 + O(1/n^{10}), \quad \text{for every } k \in [n].$$

We state the following two lemmas and use them to prove Theorem 2.

**Lemma 4.** *Let $\epsilon = n^{-6}$. Then for all $i, j : 1 \leq i \neq j \leq n$, we have*

$$(\mathbf{A}_i - \mathbf{A}_j)\mathbf{y}'^T > \epsilon/2 \implies x'_j = 0 \quad \text{and} \quad \mathbf{x}'(\mathbf{B}_i - \mathbf{B}_j) > \epsilon/2 \implies y'_j = 0, \quad (15)$$

*where $\mathbf{A}_i$ denotes the $i$th row of $\mathbf{A}$ and $\mathbf{B}_i$ denotes the $i$th column of $\mathbf{B}$.*

**Lemma 5.** *There exist $i$ and $j \in [n]$ such that*

$$x'_i \geq 1 - O(1/n^{11}) \quad \text{and} \quad y'_j \geq 1 - O(1/n^{11}).$$

Now assume that $\mathbf{x}'$ and $\mathbf{y}'$ satisfy both properties. In particular, Lemma 5 implies that $\mathbf{x}', \mathbf{y}' \neq \mathbf{0}$. Therefore, we can normalize them to get two probability distributions $\mathbf{x}$ and $\mathbf{y}$ using (4). Before proving these two lemma, we use them to show that $(\mathbf{x}, \mathbf{y})$ must be an $\epsilon$-well-supported Nash equilibrium of $(\mathbf{A}, \mathbf{B})$.

*Proof (Proof of Theorem 2).* Since both $\mathbf{x}$ and $\mathbf{y}$ are probability distributions, we only need to show that $(\mathbf{x}, \mathbf{y})$ satisfies (2) for all $i, j : 1 \leq i \neq j \leq n$. We only prove the first part of (2) here. Assume $\mathbf{A}_i \mathbf{y}^T - \mathbf{A}_j \mathbf{y}^T > \epsilon$, then

$$(\mathbf{A}_i - \mathbf{A}_j)\mathbf{y}'^T = (\mathbf{A}_i - \mathbf{A}_j)\mathbf{y}^T \cdot \sum\nolimits_{k \in [n]} y'_k > \big(1 - O(1/n^{11})\big)\epsilon > \epsilon/2,$$

by Lemma 5. As a result, by Lemma 4 we have $x'_j = 0$ and thus, $x_j = 0$. $\square$

### 5.4 Proofs of Lemma 4 and Lemma 5

*Proof (Proof of Lemma 4).* Without loss of generality, we prove the first part of (15) for the case when $i = 1, j = 2$. The other part can be proved similarly.

Let $\mathbf{u} = (1, 2, 1)$, $\mathbf{C} = \mathbf{A}_1 - \mathbf{A}_2$, $m$ be the number of non-zero entries in $\mathbf{C}$, and $C = \sum_{k \in [n]} C_k$. Assume $(\mathbf{A}_1 - \mathbf{A}_2)\mathbf{y}'^T > \epsilon/2$. Then the money of $T_\mathbf{u}$ left after purchasing the bundle in (10) is given in (11). By the definition of $y'_k$,

$$\sum_{k \in I} C_k \cdot p_{2(n+k)} = \frac{1}{3} \sum_{k \in I} C_k \cdot \big(p_{2(n+k)-1} + p_{2(n+k)}\big) + (\mathbf{A}_1 - \mathbf{A}_2)\mathbf{y}'^T.$$

By Lemma 1, $p_{2(n+k)-1} + p_{2(n+k)} \geq 3 - O(1/n^{11})$. So the money left is at most

$$\frac{3}{n^{12}} + \frac{6m + C}{n^{13}} - \frac{1}{n^{13}} \sum_{k \in I} \left(2 + \frac{C_k}{3}\right)(3 - O(1/n^{11})) - \frac{\epsilon}{2n^{13}} < \frac{3}{n^{12}} - \frac{1}{2n^{19}} + O\left(\frac{1}{n^{24}}\right).$$

After purchasing $1/n^{12}$ amount of $G_3$, even if $T_{\mathbf{u}}$ spends all the money left on $G_4$, the amount of $G_4$ she can get is at most

$$\frac{1}{p_4}\left(\frac{3}{n^{12}} - \Omega\left(\frac{1}{n^{19}}\right) - \frac{p_3}{n^{12}}\right) = \frac{1}{p_4}\left(\frac{3}{n^{12}} - \frac{p_3 + p_4}{n^{12}}\right) + \frac{1}{n^{12}} - \frac{1}{p_4} \cdot \Omega\left(\frac{1}{n^{19}}\right).$$

Since $p_3 + p_4 \geq 3 - O(1/n^{11})$, it is at most $1/n^{12} - \Omega(1/n^{19})$. By Corollary 2,

$$s_4[\mathcal{T}_U \cup \mathcal{T}_V] - a_4[\mathcal{T}_U \cup \mathcal{T}_V] \geq s_4[T_{\mathbf{u}}] - a_4[T_{\mathbf{u}}] \geq \Omega(1/n^{19}).$$

It then follows directly from Lemma 2 that $x_2' = 0$.

*Proof (Proof of Lemma 5).* Let $k$ be one of the indices that maximize $\mathbf{A}_i \mathbf{y}'^T$:

$$\mathbf{A}_k \mathbf{y}'^T = \max_i \mathbf{A}_i \mathbf{y}'^T.$$

We will show $p_{2k} = 2(p_{2k-1} + p_{2k})/3$ and then Lemma 5 follows from Lemma 1.

To this end, we bound $s_{2k}[\mathcal{T}_U \cup \mathcal{T}_V] - a_{2k}[\mathcal{T}_U \cup \mathcal{T}_V]$. By Corollary 2 we have $s_{2k}[\mathcal{T}_V] = a_{2k}[\mathcal{T}_V]$. Now let $\mathbf{u} = (i, j, 1)$ be a triple in $U$ with $i \neq j$. We consider the following two cases.

First, if $j \neq k$, then $s_{\mathbf{u},2k} = a_{\mathbf{u},2k} = 0$.

Second, if $j = k$, then we use $\mathbf{C}$ to denote $\mathbf{A}_i - \mathbf{A}_k$, $m$ to denote the number of nonzero entries in $\mathbf{C}$ and $C$ to denote $\sum_\ell C_\ell$. The way we pick $k$ guarantees that $\mathbf{C}\mathbf{y}'^T \leq 0$. After buying the bundle in (10), the money of $T_{\mathbf{u}}$ left is given in (11). By the definition of $y_k'$, it is at least

$$\frac{3}{n^{12}} + \frac{6m + C}{n^{13}} - \frac{1}{n^{13}} \sum_{k \in I} \left(2 + \frac{C_k}{3}\right)\left(3 + O(1/n^{10})\right) = \frac{3}{n^{12}} - O\left(\frac{1}{n^{23}}\right).$$

As a result, the amount of $G_{2k}$ that $T_{\mathbf{u}}$ gets is at least

$$\frac{1}{p_{2k}}\left(\frac{3}{n^{12}} - O\left(\frac{1}{n^{23}}\right) - \frac{p_{2k-1}}{n^{12}}\right) = \frac{1}{p_{2k}}\left(\frac{3}{n^{12}} - \frac{p_{2k-1} + p_{2k}}{n^{12}}\right) + \frac{1}{n^{12}} - \frac{1}{p_{2k}} \cdot O\left(\frac{1}{n^{23}}\right)$$

$$\geq 1/n^{12} - O(1/n^{22}).$$

Since the number of $\mathbf{u} \in U$ whose second component equals $k$ is $n - 1$, we have

$$s_{2k}[\mathcal{T}_U] - a_{2k}[\mathcal{T}_U] \leq O(1/n^{21}).$$

It then follows from Lemma 3 that $p_{2k} = 2(p_{2k-1} + p_{2k})/3$.

## Acknowledgement

We would like to thank Nikhil Devanur for suggesting us to consider the complexity of Fisher equilibria for additively separable PLC utility functions and for his valuable intuition on this problem.